\documentclass[aps,prc,twocolumn,preprintnumbers,amsmath,amssymb,superscriptaddress,nofootinbib]{revtex4-1}
\usepackage{amsthm}
\usepackage{mathtools}
\usepackage{physics}
\usepackage{graphicx}
\usepackage{adjustbox}
\usepackage{placeins}
\usepackage[T1]{fontenc}

\usepackage{float}

\usepackage{caption}
\usepackage{subcaption}

\usepackage{multirow}
\usepackage{graphicx}
\usepackage{dcolumn}
\usepackage{bm}
\usepackage{color}
\usepackage{threeparttable}
\usepackage{comment}
\usepackage[dvipsnames]{xcolor}
\usepackage[pdftex,colorlinks,citecolor=blue,urlcolor=blue,linkcolor=blue,bookmarks]{hyperref}

\begin{document}
\title{Muon Capture on $^{6}$Li, $^{12}$C, and $^{16}$O from \emph{Ab Initio} Nuclear Theory}

\author{Lotta Jokiniemi}
\email{ljokiniemi@triumf.ca}
\affiliation{TRIUMF, 4004 Wesbrook Mall, Vancouver, BC V6T 2A3, Canada}
\author{Petr Navr\'{a}til}
\email{navratil@triumf.ca}
\affiliation{TRIUMF, 4004 Wesbrook Mall, Vancouver, BC V6T 2A3, Canada}
\affiliation{University of Victoria, 3800 Finnerty Road, Victoria, British Columbia V8P 5C2, Canada}
\author{Jenni Kotila}
\email{jenni.kotila@jyu.fi}
\affiliation{Finnish Institute for Educational Research, University of Jyv\"askyl\"a, P.O. Box 35, Jyv\"askyl\"a FI-40014, Finland}
\affiliation{Center for Theoretical Physics, Sloane Physics Laboratory, Yale University, New Haven, Connecticut 06520-8120, USA}
\affiliation{International Center for Advanced Training and Research in Physics (CIFRA), 409, Atomistilor Street, Bucharest-Magurele, 077125, Romania}
\author{Kostas Kravvaris}
\email{kravvaris1@llnl.gov}
\affiliation{Lawrence Livermore National Laboratory, P.O. Box 808, L-414, Livermore, CA 94551, USA}
\date{\today}

\begin{abstract}
Muon capture on nuclei is one of the most promising probes of the nuclear electroweak current driving the yet-hypothetical neutrinoless double-beta ($0\nu\beta\beta$) decay. Both processes involve vector and axial-vector currents at finite momentum transfer, $q\sim 100$ MeV, as well as the induced pseudoscalar and weak-magnetism currents. Comparing measured muon-capture rates with reliable \emph{ab initio} nuclear-theory predictions could help us validate these currents.
To this end, we compute partial muon-capture rates for $^{6}$Li, $^{12}$C and $^{16}$O, feeding the ground and excited states in $^{6}$He, $^{12}$B and $^{16}$N, using \emph{ab initio} no-core shell model with two- and three-nucleon chiral interactions. We remove the spurious center-of-mass motion by introducing translationally invariant operators and approximate the effect of hadronic two-body currents by Fermi-gas model. We solve the bound-muon wave function from the Dirac wave equations in the Coulomb field created by a finite nucleus. We find that the computed rates to the low-lying states in the final nuclei are in good agreement with the measured counterparts. We highlight sensitivity of some of the transitions to the sub-leading three-nucleon interaction terms. We also compare summed rates to several tens of final states with the measured total capture rates and note that we slightly underestimate the total rate with this simple approach due to limited range of excitation energies.
\end{abstract}

\maketitle

\section{Introduction}

A negatively charged muon can replace an electron in an atomic orbit, forming a muonic atom. Since a muon is more massive than an electron, it is likely that the muon is captured by the positively charged nucleus. Ordinary muon capture (OMC), distinguished from its radiative counterpart, is a weak-interaction process in which the muon $\mu^-$ is captured by a nucleus ($A,Z$), transforming the nucleus to $(A,Z-1)^*$ and emitting a muon neutrino $\nu_{\mu}$ \cite{Measday2001}. 
The momentum transfers involved in muon capture processes, dominated by the large muon mass, are of the order of $\sim 100$ MeV. In fact, muon capture alongside neutrino-nucleus scattering are the only known nuclear-weak processes operating at this high-momentum-exchange regime. Thanks to the high momentum transfer, OMC can also lead to highly excited nuclear states with practically all spin-parities. These properties make OMC a particularly promising probe for the hypothetical neutrinoless double-beta ($0\nu\beta\beta$) decay \cite{Kortelainen2002,Kortelainen2003,Kortelainen2004,Agostini2022}. 

Muon capture operators involve both vector and axial-vector weak currents and induced magnetic and pseudoscalar currents. For a long time it was known that operators involving spin were overpredicted by nuclear theories \cite{Towner1987}, often called as ``$g_{\rm A}$ quenching'' puzzle. While a solution of the puzzle was recently proposed in the context of $\beta$ decays by introducing missing correlations and two-body currents via \textit{ab initio} nuclear theories \cite{Gysbers2019}, the situation at higher momentum exchange, relevant for $0\nu\beta\beta$ decay, is much less known. Comparing \emph{ab initio} calculations on OMC rates with experimental counterparts is one of the most promising ways to shed light on this matter.

Muon captures in light to heavy nuclei have been measured in several nuclei \cite{Measday2001}, the experiments mostly dating back to a few decades ago.
Inspired by its connection to $0\nu\beta\beta$-decay, there has been a renaissance of OMC experiments dedicated to measure OMC in nuclei involved in $0\nu\beta\beta$ processes \cite{Hashim2018,Zinatulina2019}, and these studies are planned to be extended to the remaining $\beta\beta$-decay cases as well as selected light nuclei by the MONUMENT Collaboration \cite{MON2023} at PSI, Switzerland.

Traditionally, theory predictions of OMC have been based on phenomenological models, such as the nuclear shell model (NSM) \cite{Siiskonen1998,Siiskonen1999,Kortelainen2000,Siiskonen2001,Kortelainen2002,Kortelainen2004,Suzuki2018} and proton-neutron quasiparticle random-phase approximation (pnQRPA) \cite{Giannaka2015,Jokiniemi2019b,Simkovic2020,Ciccarelli2020} frameworks. More recently, there are also \emph{ab initio} calculations for the muon-capture rates in light nuclei. Partial capture rate for the transition $\mu^-+~^{12}{\rm C}(0^+_{\rm gs})\rightarrow \nu_{\mu}+~^{12}{\rm B}(1^+_{\rm gs})$ has been calculated in no-core shell-model (NCSM) \cite{Hayes2003}. More recently, muon capture rates in $^3$He, $^4$He and $^6$Li have been computed using \emph{ab initio} Green's function Monte Carlo (GFMC) and variational Monte Carlo (VMC) methods \cite{Lovato2019,King2022}. Furthermore, partial OMC rates to low-lying states in $^{24}$Na have been evaluated in the valence-space in-medium similarity renormalization group (VS-IMSRG) method \cite{Jokiniemi2023}. These calculations provide a first step towards understanding the ``$g_{\rm 
A}$ quenching puzzle'' at finite momentum transfer.  

In the present study, we extend these \emph{ab initio} studies by computing OMC rates to the ground and low-lying excited states in $^6$He, $^{12}$B and $^{16}$N using no-core shell model (NCSM)~\cite{Navratil2000a,Navratil2000,Barrett2013} with two- and three-nucleon forces derived from the chiral effective field theory ($\chi$EFT). We use translationally invariant operators with exact bound-muon wave functions and approximate the effect of two-body currents using the Fermi-gas model. We compare the calculated rates against measured partial capture rates and earlier \emph{ab initio} calculations.

The paper is organized as follows: In Sec. \ref{sec:theory} we present the theory framework, including the bound-muon wave functions, nuclear wave functions, muon capture operators, and normal-ordered two-body currents. In Sec. \ref{sec:results} we present the obtained results for the nuclear spectroscopy and muon-capture rates. We also analyze the effect of the spurious center-of-mass motion and the dependence on the harmonic-oscillator basis. In Sec. \ref{sec:summary} we summarize the obtained results and give an outlook for future studies.

\section{Theoretical Framework}
\label{sec:theory}
\subsection{Bound-Muon Wave Functions}

\begin{figure}
    \centering
    \includegraphics{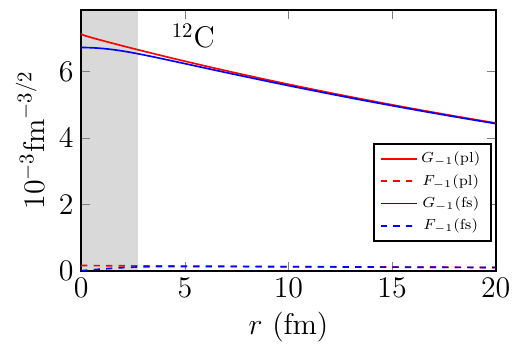}
    \caption{The large ($G_{-1}$, solid lines) and small ($F_{-1}$, dashed lines) parts of the wave function bound in the $1s_{1/2}$ atomic orbit of $^{12}$C. The blue curves show the results obtained with a finite-size (fs) nucleus, while the red curves show those obtained assuming a point-like (pl) nucleus.}
    \label{fig:muon_wave_function}
\end{figure}
The wave function of a muon bound in an atomic orbit of the mother nucleus can be expressed as an expansion in terms of the normalized spherical spinors $\chi_{\kappa\mu}$ 
\begin{equation}
\psi_{\mu}(\kappa,\mu;\mathbf{r})=\psi_{\kappa\mu}^{(\mu)}=\begin{bmatrix}-iF_{\kappa}(r)\chi_{-\kappa\mu}\\G_{\kappa}(r)\chi_{\kappa\mu}\end{bmatrix}\;,
\label{eq:muonwavefunction}
\end{equation}
where $G_{\kappa}$ and $F_{\kappa}$ are the radial wave functions of the bound state \cite{Morita1960}. Here $\kappa$ denotes the atomic orbit in the following manner
\begin{equation}
\begin{cases}
l=\kappa~\text{and}~j=l-\tfrac12,\quad \text{for}~\kappa>0\\
l=-\kappa-1~\text{and}~j=l+\tfrac12,\quad \text{for}~\kappa<0.
\end{cases}
\label{eq:kappa}
\end{equation}
Like electrons, the muon spinor corresponds to opposite $\kappa$ numbers for the large and small components of the wave function due to properties of the spherical spinors \cite{Szmytkowski2007}.

After being stopped in the outer shells of an atom, the negative muon goes trough a series of transitions to lower atomic orbitals, leaving it finally on the lowest, $K$ atomic orbit. 
Hence, the captured muon can be assumed to be initially bound in the lowest state $1s_{1/2}$, corresponding to $\kappa=-1$ and $\mu=\pm\tfrac12$.

In order to take the finite size of the nucleus properly into account, we construct a realistic bound-muon wave function by solving the Dirac wave equations~\cite{Jokiniemi2021} for the large, $G_{-1}$, and small, $F_{-1}$, parts of the wave function \eqref{eq:muonwavefunction} in the Coulomb field created by the nucleus.
Assuming the muon is bound in the lowest state $1s_{1/2}$ ($\kappa=-1$), the components satisfy the coupled differential equations
\begin{equation}
\begin{cases}
\frac{\rm d}{{\rm d}r}G_{-1}(r)+\frac{1}{r}G_{-1}(r)=\frac{1}{\hbar c}(mc^2-E+V(r))F_{-1}(r)\;,\\
\frac{\rm d}{{\rm d}r}F_{-1}(r)-\frac{1}{r}F_{-1}(r)=\frac{1}{\hbar c}(mc^2+E-V(r))G_{-1}(r)\;.
\end{cases}
\label{eq:coupled-differentials}
\end{equation}

Taking a uniform distribution of the nuclear charge within the charge radius $R_c=r_0A^{1/3}$, where $r_0=1.2$ fm, the potential energy $V(r)$ in Eqs.~\eqref{eq:coupled-differentials} can be written:
\begin{equation}
V(r)=
  \begin{cases}
-\frac{Ze^2}{2R_c}\left[3-\left(\frac{r}{R_c}\right)^2\right]\;, & \text{if $r\leq R_c$} \\
-\frac{Ze^2}{r}\;, & \text{if $r>R_c$.}
  \end{cases}
\label{eq:muonpotentialenergy}
\end{equation}
The equations \eqref{eq:coupled-differentials} can then be solved by means of the package \textsc{Radial} \cite{Salvat1995}
using a piece-wise-exact power-series expansion of the radial functions, which are summed to the prescribed accuracy. See an example of the obtained wave functions in Fig. \ref{fig:muon_wave_function}. The normalization of the obtained wave functions is defined as 
\begin{equation}
N=\int [G_{-1}^2(r)+ F_{-1}^2(r)]r^2dr=1.
\label{eq:muonnormalization}
\end{equation}
A similar method has previously been used for both bound and scattering electron wave functions in the context of $\beta\beta$ decay \cite{Kotila2013, Kotila2012}, and for muon capture \cite{Jokiniemi2021,Jokiniemi2023}.

\subsection{Nuclear Wave Functions and Transition Operators}

Nuclear wave functions and the transition operator matrix elements are obtained within the NCSM. In this approach, nuclei are considered to be systems of $A$ non-relativistic point-like nucleons interacting via realistic chiral two-nucleon (NN) and three-nucleon (3N) interactions. Each nucleon is an active degree of freedom and the total momentum,
the angular momentum, and the parity of the nucleus are conserved. The many-body wave function is expanded over a basis of antisymmetric $A$-nucleon harmonic oscillator (HO) states. The basis contains up to $N_{\rm max}$ HO excitations above the lowest possible Pauli configuration. The basis is characterized by an additional parameter $\Omega$, the frequency of the HO well. The convergence of the HO expansion can be greatly accelerated by applying a Similarity Renormalization Group (SRG) transformation on the 2N and 3N interactions~\cite{Bogner2007}. 

We use three different $\chi$EFT NN and 3N interactions: NN-N$^3$LO \cite{Entem2003}+3N$_{\rm lnl}$ \cite{Soma2020}, NN-N$^4$LO \cite{Entem2017}+3N$_{\rm lnl}$ \cite{Gysbers2019}, and NN-N$^4$LO+3N$^*_{\rm lnl}$~\cite{Kravvaris2023}, where an additional sub-leading contact term ($E_7$) enhancing the spin-orbit strength~\cite{Girlanda2011} has been introduced to the 3N force. The $E_7$ low-energy constant has been adjusted to improve the description of the excitation energies of $^6$Li, in particular of the first excited state $3^+$, $T=0$. Both high-precision NN-N$^3$LO and NN-N$^4$LO interactions use a 500 MeV regulator cutoff. The interactions have been softened by the Similarity Renormalization Group (SRG) technique \cite{Bogner2007} with the SRG induced three-nucleon terms fully included. In the present study, we use the evolution parameter $\lambda_{\rm SRG}=2.0$ fm$^{-1}$ for the NN-N$^3$LO+3N$_{\rm lnl}$ and NN-N$^4$LO+3N$^*_{\rm lnl}$ interactions, and $\lambda_{\rm SRG}=1.8$ fm$^{-1}$ for the NN-N$^4$LO+3N$_{\rm lnl}$ interaction. We have checked that the observables are insensitive to the variation of the $\lambda_{\rm SRG}$ parameter between 1.8 and 2.0 fm$^{-1}$.

In the lightest systems, $^{6}$Li and $^{6}$He, we are able to reach large model spaces up to $N_{\rm max}=14$ without additional truncations. However, for the $N_{\rm max}=8$ calculations in the $A=12,16$ systems, additional truncation is needed in order to make the calculations feasible. To that end, we use importance truncation \cite{Roth2007,Roth2009} to control the basis size.

We define the reduced nuclear matrix element (NME) of an operator $\mathcal{O}_s$ for a transition from an initial state $\Psi_{J_iM_i}$ to a final state $\Psi_{J_fM_f}$ as 
\begin{equation}
\begin{split}
\int &\Psi_{J_fM_f}\sum_{s=1}^{A}\mathcal{O}_s\tau^s_-\Psi_{J_iM_i}
d\mathbf{r}_1...d\mathbf{r}_A\\
=&\frac{1}{\sqrt{2J_f+1}}\langle\Psi_f||\sum_{s=1}^A\mathcal{O}_{s}(\mathbf{r}_s,\mathbf{p}_s)\tau_-^s||\Psi_i\rangle\\
&\times(J_iM_iuM_f-M_i|J_fM_f)\;,
\end{split}
\label{eq:ME-definition}
\end{equation}
where $\tau_-^s$ is the isospin lowering operator changing a neutron into a proton and $u$ is the rank of the operator $\mathcal{O}_s$ and the summation runs over all the $A$ nucleons. 
The different operators $\mathcal{O}_{s}$ contributing to muon-capture rates are defined in Table \ref{tab:operators}. We use the same notation as in Ref. \cite{Morita1960} and denote the reduced matrix elements corresponding to different operators as $\mathcal{M}[k\,w\,u\,\binom{\pm}{p}]$, where $k$, $w$ and $u$ are the spin, orbital and the total angular momenta (or the rank) of the operator $\mathcal{O}_s$ and $+$, $-$ and $p$ are additional symbols referring to derivative and gradient operators (see Table \ref{tab:operators} and Ref. \cite{Morita1960} for more details). 
Here we assume that the muon is bound on the $\kappa=-1$ orbit and that the small component of the bound-muon wave function is negligible.

{ 
\renewcommand{\arraystretch}{2.0}
\begin{table*}[t]\centering
\caption{Definition of $\mathcal{O}_{s}$ in Eq. \eqref{eq:ME-definition} for different OMC nuclear matrix elements (NMEs).
}
\begin{ruledtabular} 
\begin{tabular}{ll}
\centering NME & $\mathcal{O}_s$\\
\hline
$\mathcal{M}[0\,w\,u]$ &$j_w(qr_s)G_{-1}(r_s)\mathcal{Y}_{0wu}^{M_f-M_i}(\hat{\mathbf{r}}_s)\delta_{wu}$\\
$\mathcal{M}[1\,w\,u]$ &$j_w(qr_s)G_{-1}(r_s)
\mathcal{Y}_{1wu}^{M_f-M_i}(\hat{\mathbf{r}}_s,\boldsymbol{\sigma}_s)$\\
$\mathcal{M}[0\,w\,u\,\pm]$ &$[j_w(qr_s)G_{-1}(r_s)\mp\frac{1}{q} j_{w\mp 1}(qr_s)\frac{d}{dr_s}G_{-1}(r_s)]
\mathcal{Y}_{0wu}^{M_f-M_i}(\hat{\mathbf{r}}_s)\delta_{wu}$\\
$\mathcal{M}[1\,w\,u\,\pm]$ &$[j_w(qr_s)G_{-1}(r_s)\mp\frac{1}{q} j_{w\mp 1}(qr_s)\frac{d}{dr_s}G_{-1}(r_s)]
\mathcal{Y}_{1wu}^{M_f-M_i}(\hat{\mathbf{r}}_s,\boldsymbol{\sigma}_s)$\\
$\mathcal{M}[0\,w\,u\,p]$ &$ij_w(qr_s)G_{-1}(r_s)\mathcal{Y}_{0wu}^{M_f-M_i}(\hat{\mathbf{r}}_s)
\boldsymbol{\sigma}_s\cdot \mathbf{p}_s \delta_{wu}$\\
$\mathcal{M}[1\,w\,u\,p]$ &$ij_w(qr_s)G_{-1}(r_s)
\mathcal{Y}_{1wu}^{M_f-M_i}(\hat{\mathbf{r}}_s,\mathbf{p}_s)$\\
\end{tabular}
\end{ruledtabular}
\label{tab:operators}
\end{table*}
}

In Table~\ref{tab:operators}, $j_w(qr_s)$ is the spherical Bessel function of rank $w$ and the quantities $\mathcal{Y}_{kwu}^M$ are the (vector) spherical harmonics defined as
\begin{equation}
\begin{split}
\mathcal{Y}_{0wu}^M(\hat{\mathbf{r}})\equiv & (4\pi)^{-1/2}Y_{w,M}(\hat{\mathbf{r}})\;,\\
\mathcal{Y}_{1wu}^M(\hat{\mathbf{r}},\boldsymbol{\sigma})\equiv
&\sum_m(1\ -m\ w\ m+M|u\,M)\\
&\times Y_{w,m+M}(\hat{\mathbf{r}})\sqrt{\frac{3}{4\pi}}\sigma_{-m}\;,\\
\mathcal{Y}_{1wu}^M(\hat{\mathbf{r}},\mathbf{p})\equiv
&\sum_m(1\ -m\ w\ m+M|u\,M)\\
&\times Y_{w,m+M}(\hat{\mathbf{r}})\sqrt{\frac{3}{4\pi}}p_{-m}\;,\\
\end{split}
\label{eq:spherical-harmonics}
\end{equation}
where $\boldsymbol{\sigma}$ is the Pauli spin vector, $\mathbf{p}$ is the nucleon momentum, $Y_{w,M}(\hat{\mathbf{r}})$ are the spherical harmonics, and $\hat{\mathbf{r}}$ is the unit coordinate vector for angles in spherical coordinates.

The $Q$-value of the muon-capture process is obtained as
\begin{equation}
q=(m_{\mu}-W_0)\left(1-\frac{m_{\mu}}{2(m_{\mu}+AM)}\right)\;,
    \label{eq:q}
\end{equation}
where $W_0=M_f-M_i+m_e+E_X$. 
Here $M_f$($M_i$) is the nuclear mass of the final(initial) nucleus, $m_e$ the rest mass of an electron, $m_{\mu}$ the rest mass of a muon, $M$ the average nucleon mass, and $E_X$ the excitation energy of the final $J^{\pi}$ state. 

We compute the reduced matrix elements of the one-body operators between initial- and final-state NCSM wave functions by introducing charge-changing one-body transition densities as
\begin{equation}
\begin{split}
   \langle\Psi_f||\sum_{s=1}^A\mathcal{O}_{s}&(\mathbf{r}_s,\mathbf{p}_s)\tau_-^s||\Psi_i\rangle=-\frac{1}{\sqrt{2u+1}}\\ &\times\sum_{\pi\nu}\langle\nu||\mathcal{O}_{s}(\mathbf{r}_s,\mathbf{p}_s)\tau_-^s||\pi\rangle\\
    &\times\langle\Psi_f||[a^{\dag}_{\nu}\Tilde{a}_{\pi}]_u||\Psi_i\rangle\;,
    \end{split}
\end{equation}
where $\mathcal{O}_{s}$ are the one-body operators from Table \ref{tab:operators}, $\pi$ and $\nu$ label the different proton and neutron orbitals and $\Tilde{a}_{\pi,m_{\pi}}=(-1)^{j_{\pi}-m_{\pi}}a_{{\pi},-m_{\pi}}$ with $a^{\dag}_{\alpha}$ and $a_{\beta}$ the creation and annihilation operators of the HO single-particle states.

The single-particle coordinates $\mathbf{r}_s$ in the definitions of the operators in Tab. \ref{tab:operators} are measured with the respect to the center of mass of the harmonic oscillator potential. To remove the spurious center-of-mass motion caused by this, we follow Ref. \cite{Navratil2004} and introduce translationally invariant one-body densities depending on coordinates and momenta measured from the center of mass of the nucleus, i.e. $\boldsymbol{\xi}_s=-\sqrt{A/(A-1)}(\mathbf{r}_s-\mathbf{R}_{\rm CM})$ and $\boldsymbol{\varpi}_s=-\sqrt{A/(A-1)}(\mathbf{p}_s-\mathbf{P})$, where $\mathbf{R}_{\rm CM}=\tfrac{1}{A}\sum_{s=1}^A\mathbf{r}_s$ and $\mathbf{P}=\sum_{s=1}^A\mathbf{p}_s$ are the center-of-mass coordinate and momentum of the $A$-nucleon system. We can then obtain translationally invariant matrix elements for the operators following Ref. \cite{Navratil2021} as
\begin{equation}
\begin{split}
\langle\Psi_f||\sum_{s=1}^A&\mathcal{O}_{s}(\mathbf{r}_s-\mathbf{R}_{\rm CM},\mathbf{p}_s-\mathbf{P})\tau_-^s||\Psi_i\rangle=-\frac{1}{\sqrt{2u+1}}\\ &\times\sum_{\pi\nu\pi'\nu'}\langle{\nu'}||\mathcal{O}_{s}\left(-\sqrt{\tfrac{A-1}{A}}\boldsymbol{\xi}_s,-\sqrt{\tfrac{A-1}{A}}\boldsymbol{\varpi}_s\right)\tau_-^s||{\pi'}\rangle\\
    &\times(M^u)^{-1}_{{\nu'\pi'},{\nu\pi}}\langle\Psi_f||[a^{\dag}_{\nu}\Tilde{a}_{\pi}]_u||\Psi_i\rangle\;,
    \end{split}
\end{equation}
where $M^u$ is the transformation matrix defined in \cite{Navratil2021}.
\subsection{Capture Rates}

Following Ref. \cite{Morita1960}, the capture rate for a transition from a $J_i^{\pi}$ initial state to a $J_f^{\pi}$ final state can be written as \footnote{Our definition of the capture rate differs from that in Ref. \cite{Morita1960} by a factor of $1/(2J_f+1)$ because of our definition of the reduced matrix elements \eqref{eq:ME-definition} consistent with \cite{Edmonds1960}.} 
\begin{equation}
  W=2P\frac{1}{2J_i+1}\left(1-\frac{q}{m_{\mu}+AM}\right)q^2 \;,
\end{equation}
where the term $P$ contains the reduced NMEs and can be written in terms of vector (V), axial-vector (A), magnetic (M) and pseudoscalar (P) contributions as
\begin{equation}
\label{eq:p}
P=\frac{1}{2}\sum_{\kappa' u}\Big|M^{\rm V}_{\kappa' u}+M^{\rm A}_{\kappa' u}+M^{\rm M}_{\kappa' u}+M^{\rm P}_{\kappa' u}\Big|^2\;,
\end{equation}
where the summation goes over the neutrino quantum numbers $l'$ and $j'$, abbreviated as $\kappa'$ as in Eq. \eqref{eq:kappa}, and the operator rank $u$ is restricted by $|J_i-J_f|\leq u\leq J_i+J_f$.

The vector part in Eq. \eqref{eq:p} is defined as
\begin{equation}
\begin{split}
M^{\rm V}_{\kappa' u}=&g_{\rm V}(q^2)\bigg[\mathcal{M}[0\,l'\,u]S_{0u}(\kappa')\delta_{l'u}\\
&-\frac{1}{M}
\mathcal{M}[1\,\bar{l'}\,u\,p]S'_{1u}(-\kappa')\\ &+\sqrt{3}\frac{q}{2M}\bigg(\sqrt{\frac{\bar{l'}+1}{2\bar{l'}+3}}
\mathcal{M}[0\,\bar{l'}\!+\!1\,u\,+]\delta_{\bar{l'}+1,u}\\
 &+ \sqrt{\frac{\bar{l'}}{2\bar{l'}-1}}\mathcal{M}[0\,\bar{l'}\!-\!1\,u\,-]\delta_{\bar{l'}-1,u}\bigg)S'_{1u}(-\kappa')\bigg]\;,
 \end{split}
\label{eq:M_V}
\end{equation}
the axial-vector part as
\begin{equation}
    \begin{split}
M^{\rm A}_{\kappa' u}=&g_{\rm A}(q^2)\bigg[-\mathcal{M}[1\,l'\,u]S_{1u}(\kappa')\\
&-\frac{1}{M}\mathcal{M}[0\,\bar{l'}\,u\,p]S'_{0u}(-\kappa')\delta_{\bar{l'}u}
\\
&+\sqrt{\frac{1}{3}}\frac{q}{2M}\times\bigg(\sqrt{\frac{\bar{l'}+1}{2\bar{l'}+1}}\mathcal{M}[1\,\bar{l'}\!+\!1\,u\,+]\\
&+
\sqrt{\frac{\bar{l'}}{2\bar{l'}+1}}\mathcal{M}[1\,\bar{l'}\!-\!1\,u\,-]\bigg)\bigg] S'_{0u}(-\kappa)\delta_{\bar{l'}u}\;,
\label{eq:M_A}
    \end{split}
\end{equation}
the magnetic part as
\begin{equation}
    \begin{split}
M^{\rm M}_{\kappa' u}=&\sqrt{\frac{3}{2}}\frac{g_{\rm M}(q^2)q}{M}S'_{1u}(-\kappa')\\
&\times\Big(\sqrt{\bar{l'}+1}W(1\,1\,u\,\bar{l'}\,;1\,\bar{l'}+1)
\mathcal{M}[1\,\bar{l'}\!+\!1\,u\,+]\\
&+\sqrt{\bar{l'}}W(1\,1\,u\,\bar{l'}\,;1\,\bar{l'}-1)\mathcal{M}[1\,\bar{l'}\!-\!1\,u\,-]\Big)\;,
    \end{split}
    \label{eq:M_M}
\end{equation}
and the pseudoscalar part as
\begin{equation}
    \begin{split}
M^{\rm P}_{\kappa' u}=&-\sqrt{\frac{1}{3}}\frac{q}{2M}g_{\rm P}(q^2) S'_{0u}(-\kappa)\delta_{\bar{l'}u}\\
&\times\bigg(\sqrt{\frac{\bar{l'}+1}{2\bar{l'}+1}}\mathcal{M}[1\,\bar{l'}\!+\!1\,u\,+]\\
&+\sqrt{\frac{\bar{l'}}{2\bar{l'}+1}}\mathcal{M}[1\,\bar{l'}\!-\!1\,u\,-]\bigg)\;.
\label{eq:M_P}
    \end{split}
\end{equation}

We use the usual dipole form factors $g_{\rm A}(q^2)$ and $g_{\rm V}(q^2)$ for the axial-vector and vector couplings. For the induced weak-magnetism coupling we use $g_{\rm M}(q^2)=(1+\mu_p-\mu_n)g_{\rm V}(q^2)$ and for the pseudoscalar coupling the Goldberger-Treiman partially conserved axial-vector-current (PCAC) value
\begin{equation}
g_{\rm P}(q^2)=\frac{2Mq}{q^2+m_{\pi}^2}g_{\rm A}(q^2)\;.
\end{equation}

The $W(...)$ in Eqs. (\ref{eq:M_V}-\ref{eq:M_P}) are the usual Racah coefficients and the $S$'s are geometric factors defined as
\begin{equation}
S_{ku}(\kappa')=
\begin{cases}
\sqrt{2(2j'+1)}W(\tfrac12\, 1\,j'\,l'\,;\tfrac12 \,u)\delta_{lw}\text{ , for } k=1\\
\sqrt{\frac{2j'+1}{2l'+1}}\delta_{l'w}\text{ , for } k=0
\end{cases}
\end{equation}
and
\begin{equation}
S'_{ku}(-\kappa')={\rm sgn}(\kappa')S_{ku}(-\kappa')\;,
\end{equation}
where ${\rm sgn}(\kappa)$ is the sign of $\kappa$.
The angular momenta $l$ and $\bar{l}$ correspond to $\kappa$ and $-\kappa$ respectively. 

\subsection{Two-Body Currents}

We use two-body currents from $\chi$EFT approximated as effective one-body operators via normal ordering with respect to a spin-isospin symmetric Fermi gas reference state as in Ref.~\cite{Hoferichter2020}. The resulting current is
\begin{equation}
    \mathbf{J}^{\rm eff}_{i,\rm 2b}(\rho,\mathbf{q})=g_A\tau^-_i\bigg[\delta_ a(q^2)\bm{\sigma}_i+\frac{\delta _a^P(q^2)}{q^2}(\mathbf{q}\cdot\bm{\sigma}_i)\mathbf{q}\bigg],
\end{equation}
with two-body functions $\delta_a(q^2)$, $\delta_a^P(q^2)$ dependent on the Fermi-gas density $\rho$:
\begin{equation}
\begin{split}
\delta_a(q^2)=&-\frac{\rho}{F_{\pi}^2}\bigg[\frac{c_4}{3}[3I^{\sigma}_2(\rho,q)-I^{\sigma}_1(\rho,q)]\\
&-\frac13\bigg(c_3-\frac{1}{4m_{\rm N}}\bigg)I^{\sigma}_1(\rho,q)\\
&-\frac{c_6}{12}I_{c6}(\rho,q)-\frac{c_D}{4g_A\Lambda_{\chi}}\bigg]\;,
\end{split}
    \label{eq:delta_a}
\end{equation}
and
\begin{equation}
\begin{split}
\delta_a^P(q^2)=&\frac{\rho}{F_{\pi}^2}\bigg[-2(c_3-2c_1)\frac{m_{\pi}^2q^2}{(m_{\pi}^2+q^2)^2}\\
&+\frac13\bigg(c_3+c_4-\frac{1}{4m_{\rm N}}\bigg)I^P(\rho,q)\\
&-\bigg(\frac{c_6}{12}-\frac23\frac{c_1m_{\pi}^2}{m_{\pi}^2+q^2}\bigg)I_{c6}(\rho,q)\\
&-\frac{q^2}{m_{\pi}^2+q^2}\bigg(\frac{c_3}{3}[I^{\sigma}_1(\rho,q)+I^P(\rho,q)]\\
&+\frac{c_4}{3}[I^{\sigma}_1(\rho,q)+I^P(\rho,q)-3I^{\sigma}_2(\rho,q)]\bigg)\\
&-\frac{c_D}{4g_A\Lambda_{\chi}}\frac{q^2}{m_{\pi}^2+q^2}\bigg]\;.
\end{split}
    \label{eq:delta_a^P}
\end{equation}
The integrals $I^{\sigma}_1(\rho,q)$, $I^{\sigma}_2(\rho,q)$, $I_{c6}(\rho,q)$ and $I^P(\rho,q)$ are given in Ref.~\cite{Klos2015}. Following Ref.~\cite{Hoferichter2020}, we use $F_{\pi}=92.28~{\rm MeV}$ and $\Lambda_{\chi}=700~{\rm MeV}$. The low-energy constants $c_1$, $c_3$, $c_4$ and $c_D$ for each interaction are listed in Table \ref{tab:LECs} -- $c_6$ is taken from Ref.~\cite{Hoferichter2020}. 

\begin{table}[h]
    \centering
    \caption{The low-energy constants corresponding to each employed chiral interaction. For NN-N$^4$LO+3N$^*_{\rm lnl}$, the displayed low-energy constants are the same as for NN-N$^4$LO+3N$_{\rm lnl}$.}
    \begin{ruledtabular}
    \begin{tabular}{lccccc}
         Interaction &$c_1$ &$c_3$ &$c_4$ &$c_D$ &Ref.\\
         \hline
         NN-N$^4$LO+3N$_{\rm lnl}$ &-1.10(3) &-5.54(6) &4.17(4) &-1.8 &\cite{Entem2017,Gysbers2019} \\
         NN-N$^3$LO+3N$_{\rm lnl}$ &-0.81 &-3.20 &5.4 &0.7 &\cite{Entem2003,Soma2020}
    \end{tabular}
    \end{ruledtabular}
    \label{tab:LECs}
\end{table}

\begin{table}[h]
    \centering
    \caption{The effect of axial-vector 2BCs at $q=0$ MeV corresponding to NCSM studies of \cite{Gysbers2019} together with the Fermi-gas densities $\rho$ adjusted so that Eq. \eqref{eq:delta_a} gives the correct effect at $q=0$ MeV.}
    \begin{ruledtabular}
    \begin{tabular}{llcc}
         Mass number &Interaction &$\delta_a(0)$ &$\rho({\rm fm}^{-3})$\\
         \hline
         6& NN-N$^4$LO+3N$_{\rm lnl}$ &$-0.010$ &0.0075\\
         6& NN-N$^3$LO+3N$_{\rm lnl}$ &$-0.007$ &0.0085\\
         $12-16$& NN-N$^4$LO+3N$_{\rm lnl}$ &$-0.043$ &0.021\\
         $12-16$& NN-N$^3$LO+3N$_{\rm lnl}$ &$-0.040$ &0.027\\
    \end{tabular}
    \end{ruledtabular}
    \label{tab:rhos}
\end{table}

In practise, we implement the effect of the two-body currents on the muon-capture rates by replacing
$$g_{\rm A}(q^2,{\rm 2b})\rightarrow g_{\rm A}(q^2)+g_{\rm A}\delta_a(q^2)$$ and $$g_{\rm P}(q^2,{\rm 2b})\rightarrow g_{\rm P}(q^2)-\frac{2m_{\rm N}g_{\rm A}}{q}\delta_a^P(q^2)$$
in Eqs. \eqref{eq:M_A} and \eqref{eq:M_P}.

\begin{figure}
    \centering
    \includegraphics{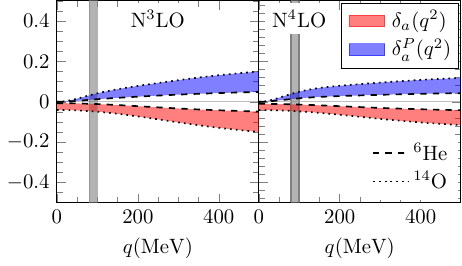}
    \caption{Two-body currents obtained from Eqs. \eqref{eq:delta_a} (red) and \eqref{eq:delta_a^P} (blue) with $\rho$ adjusted to the exact two-body currents in beta decays of $^6$He (dashed line) or $^{14}$O (dotted line). The momentum-exchange region typical in muon-capture processes is denoted by the gray band.}
    \label{fig:two-body_currents}
\end{figure}

\begin{figure*}
\includegraphics[width=\linewidth]{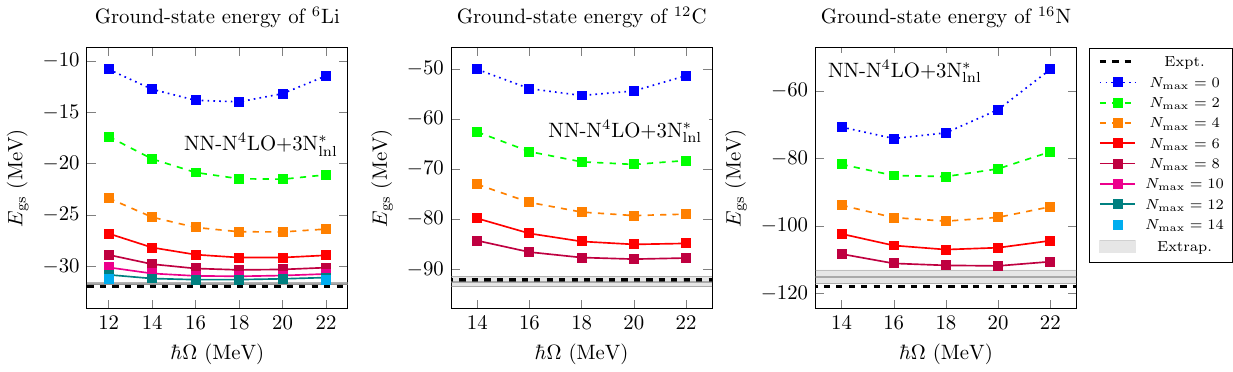}
    \caption{Ground-state energies of $^6$Li, $^{12}$C, and $^{16}$N obtained with the NN-N$^4$LO+3N$^*_{\rm lnl}$ interaction with different HO frequencies.}
    \label{fig:ground_state_energies}
\end{figure*}

As was shown in \cite{Gysbers2019}, the Fermi-gas model (in \cite{Gysbers2019}, however, the expressions for the currents are less complete than in the present work) satisfyingly captures the effect of two-body currents in $^{100}$Sn at low momentum transfer, with a certain choice of the Fermi-gas density $\rho$. We follow a similar method and adjust the Fermi-gas density $\rho$ so that Eq. \eqref{eq:delta_a} corresponds to the effect of the explicit two-body currents in the $\beta$ decays $^{6}{\rm He}~\rightarrow~^{6}{\rm Li}$ and $^{14}{\rm O}~\rightarrow~^{14}{\rm N}$ in the NCSM calculations of \cite{Gysbers2019}. The resulting $\delta_a(0)$ and $\rho$ values are listed in Table \ref{tab:rhos}, and the corresponding momentum-dependent two-body functions are shown in Fig. \ref{fig:two-body_currents}. We then use these two-body-current estimates for the muon capture on $^{6}$Li, and $^{12}$C and $^{16}$O, correspondingly. It is worthwhile to note that the model has not been validated at high momentum transfer. Furthermore, the Fermi-gas approximation might not be appropriate for light nuclei such as those considered in the present study. To properly take the effect of the two-body currents into account, we should include exact two-body currents at finite momentum transfer into our calculations, like has been done at the zero-momentum transfer limit for $\beta$ decays \cite{Gysbers2019}, magnetic dipole decays \cite{Friman-Gayer2021}, and for magnetic dipole moments \cite{Miyagi2023}. However, in the present study we do not have access to such currents and we use the adjusted Fermi-gas model as an estimate of the effect of two-body currents in the studied muon-capture transitions.

\section{Results}
\label{sec:results}
\subsection{Spectroscopy}

To test the validity of the nuclear wave functions obtained from the no-core shell model, we first explore the spectroscopic properties of the involved nuclei. The ground-state energies of the nuclei of interest are shown in Fig. \ref{fig:ground_state_energies}. The energies are obtained with  the NN-N$^4$LO+3N$^*_{\rm lnl}$ interaction with different HO frequencies. The colors denote different $N_{\rm max}$ values, and the dashed black lines show the experimental values \cite{AME2020}. We see that increasing the basis size flattens out the $\hbar\Omega$-dependence and the results converge satisfyingly towards the measured energies. to extrapolate the final results, we fit an exponential $f(N_{\rm max})=a+be^{-cN_{\rm max}}$ to the energies obtained with the frequency corresponding to the minimum of the ground-state energy and estimate the uncertainty by varying the points included in the fit. The extrapolated energies are shown in the figure as horizontal gray bands and collected in Table \ref{tab:gs_energies}. Apparently, the NN-N$^4$LO+3N$^*_{\rm lnl}$ with the sub-leading 3N term gives slightly closer binding energies to experiment compared to the other two interactions.

\begin{table}[h]
    \centering
    \caption{The extrapolated ground-state energies of $^6$He, $^6$Li, $^{12}$C, $^{12}$B, $^{16}$O and $^{16}$N.}
    \label{tab:gs_energies}
    \begin{ruledtabular}
    \begin{tabular}{llcc}
Nucleus &Interaction &$E_{\rm gs}$ (MeV) &Expt. (MeV) \cite{AME2020}\\
\hline
$\rm ^6He$ &NN-N$^3$LO+3N$_{\rm lnl}$ &-28.3(1) &-29.27\\
&NN-N$^4$LO+3N$_{\rm lnl}$&-28.3(1)\\
 &NN-N$^4$LO+3N$^*_{\rm lnl}$&-28.6(1)\\
 \hline
$\rm ^6Li$ &NN-N$^3$LO+3N$_{\rm lnl}$ &-31.5(1) &-31.99 \\
&NN-N$^4$LO+3N$_{\rm lnl}$&-31.4(1)\\
&NN-N$^4$LO+3N$^*_{\rm lnl}$ &-31.7(1) \\
\hline
$\rm ^{12}C$ &NN-N$^3$LO+3N$_{\rm lnl}$  &-88.7(10) &-92.16\\
&NN-N$^4$LO+3N$_{\rm lnl}$&-88.6(10)\\
&NN-N$^4$LO+3N$^*_{\rm lnl}$ &-92.5(10) \\ 
\hline
$\rm ^{12}B$ &NN-N$^3$LO+3N$_{\rm lnl}$  &-76.1(10) &-79.57\\ 
&NN-N$^4$LO+3N$_{\rm lnl}$&-76.0(10)\\
&NN-N$^4$LO+3N$^*_{\rm lnl}$ &-79.5(10)\\
\hline
$\rm ^{16}O$  &NN-N$^3$LO+3N$_{\rm lnl}$ & -126(2) &-127.62\\ 
&NN-N$^4$LO+3N$_{\rm lnl}$ &-127(2)\\
&NN-N$^4$LO+3N$^*_{\rm lnl}$ &-127(2)\\
\hline
$\rm ^{16}N$  &NN-N$^3$LO+3N$_{\rm lnl}$ &-114(2) &-116.58\\ 
&NN-N$^4$LO+3N$_{\rm lnl}$ &-114(2)\\
&NN-N$^4$LO+3N$^*_{\rm lnl}$ &-115(2)\\
\end{tabular}
\end{ruledtabular}
\end{table}

\begin{figure*}
    \begin{subfigure}[b]{0.45\linewidth}
    \includegraphics[width=\linewidth]{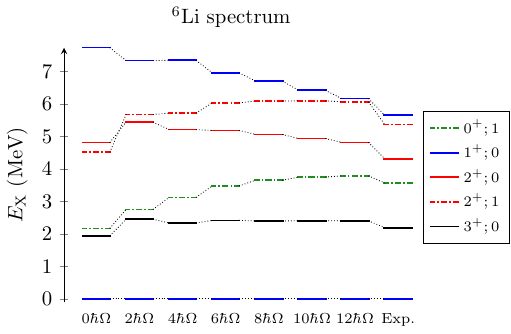}
    \end{subfigure}
    \begin{subfigure}[b]{0.45\linewidth}
    \includegraphics[width=\linewidth]{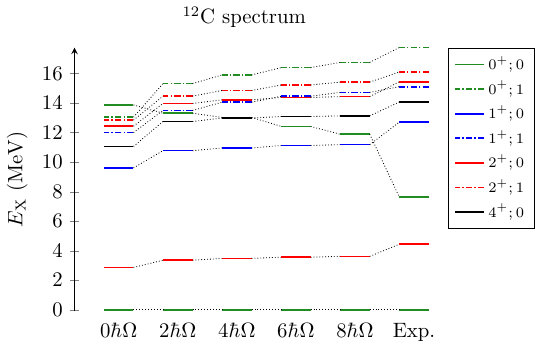}
    \end{subfigure}
    \begin{subfigure}[b]{0.45\linewidth}
    \includegraphics[width=\linewidth]{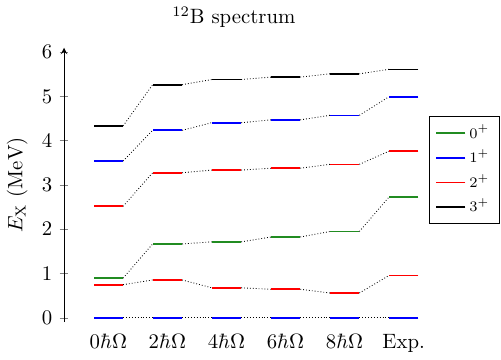}
    \end{subfigure}
    \begin{subfigure}[b]{0.45\linewidth}
    \includegraphics[width=\linewidth]{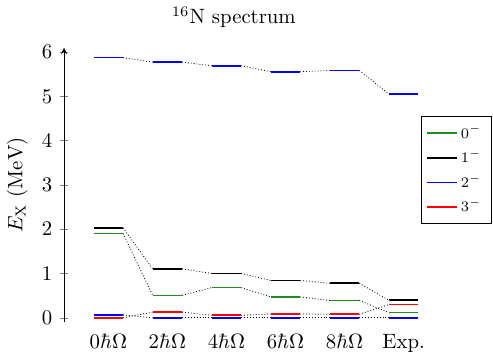}
    \end{subfigure}
    \caption{Low-energy spectrum for $^{6}$Li, $^{12}$C, $^{12}$B, and $^{16}$N (right). The theoretical values for $^{6}$Li, $^{12}$C, and $^{12}$B are obtained with the NN-N$^4$LO+3N$^*_{\rm lnl}$ and for $^{16}$N with the NN-N$^3$LO+3N$_{\rm lnl}$ interaction, all with $\hbar\Omega$=14 MeV.}
    \label{fig:spectra}
\end{figure*}

In Fig. \ref{fig:spectra}, we show the excitation-energy spectra of $^6$Li, $^{12}$C, $^{12}$B and $^{16}$N. The spectra of $^{6}$Li, $^{12}$C, and $^{12}$B are obtained with the NN-N$^4$LO+3N$^*_{\rm lnl}$ and for $^{16}$N with the NN-N$^3$LO+3N$_{\rm lnl}$ interaction, all with $\hbar\Omega$=14 MeV. The horizontal axes show increasing basis size as $N_{\rm max}\hbar\Omega$, and the computed values are compared against the measured ones \cite{TUNL}. In the spectra of $^6$Li and $^{12}$C the isospin $T=0$ states are shown as solid lines and the $T=1$ states as dash-dotted lines. The spectra of 
$^{12}$B and $^{16}$N only contain $T=1$ states. We see that with increasing basis size the energy levels are converging towards the experimental ones in all the nuclei. There are, however, a couple of exceptions: In $^{12}$C the second $0^+$ is the Hoyle state, which is known to be difficult to describe with the NCSM. In $^{16}$N, the ordering of the low-lying $0^-$ and $3^-$ states is flipped, but overall NCSM succeeds in predicting the energy levels satisfactorily. 

\subsection{Removing Spurious Center-of-Mass Motion}

\begin{figure*}
    \centering
    \begin{subfigure}[b]{0.42\linewidth}
       \includegraphics[width=\linewidth]{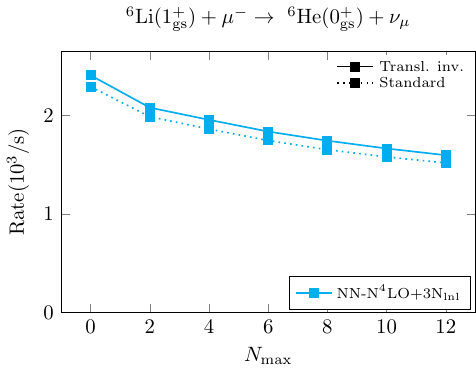} 
    \end{subfigure}
    \hspace{0.05\linewidth}
    \begin{subfigure}[b]{0.42\linewidth}
    \includegraphics[width=\linewidth]{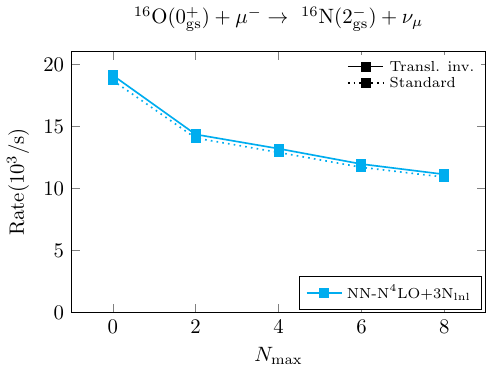}
    \end{subfigure}
      \caption{The obtained capture rates to the ground states of $^{6}$Li, $^{12}$B and $^{16}$N obtained with translationally invariant (solid lines) or standard (dotted lines) one-body densities. The rates are obtained with the NN-N$^4$LO+3N$_{\rm lnl}$ interaction with $\hbar\Omega=20$ MeV.}
    \label{fig:spurious_cms_motion}
\end{figure*}
In Fig. \ref{fig:spurious_cms_motion}, we show the effect of removing the spurious center-of-mass motion from the matrix elements of the ground-state-to-ground-state transition rates for ${\rm ^6Li}\rightarrow {\rm^6He}$ and ${\rm ^{16}O}\rightarrow {\rm ^{16}N}$. The rates computed with the standard one-body densities are shown as dotted lines while those obtained with the translationally invariant densities are shown as solid lines. We note that the effect is notable, about 4\%, for the $A=6$ system, while it is less significant, $\sim 2\%$, for the heavier systems. Overall, the effect is larger than in a previous $\beta$-decay study \cite{Glick-Magid2022}, where the same approach was followed to remove the spurious center-of-mass motion from the operators. This can be explained by the high momentum exchange involved in the muon captures. As was noted in Ref. \cite{Glick-Magid2022}, the spurious center-of-mass contamination of the operators increases with an increasing $q$. Similarly as in Ref. \cite{Glick-Magid2022}, the center-of-mass effect is largest, about $5\%-10\%$ in the case of $^6$Li, for the NMEs of the type $\mathcal{M}[kwup]$ involving the gradient operator $\mathbf{p}_s$. For the rest of the NMEs the effect is of the order $1\%-2\%$.

\subsection{Dependence on the harmonic-oscillator frequency}

\begin{figure*}
    \centering
    \begin{subfigure}[b]{0.32\linewidth}
    \includegraphics[height=0.9\linewidth]{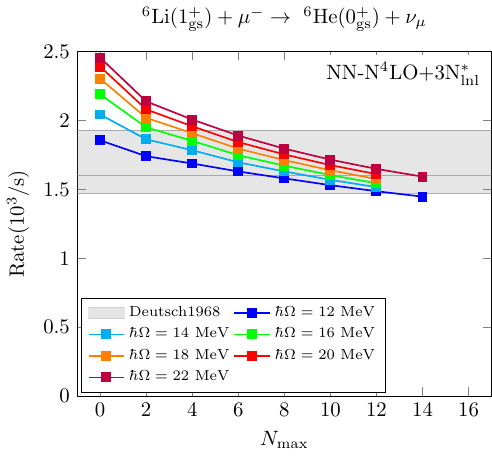}
    \end{subfigure}
    \begin{subfigure}[b]{0.32\linewidth}
    \includegraphics[height=0.9\linewidth]{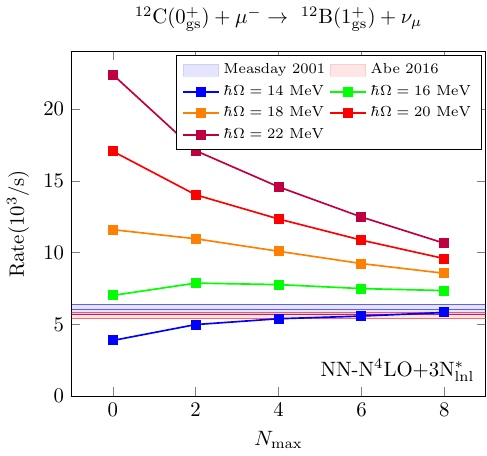}
    \end{subfigure}
    \begin{subfigure}[b]{0.32\linewidth}
    \includegraphics[height=0.9\linewidth]{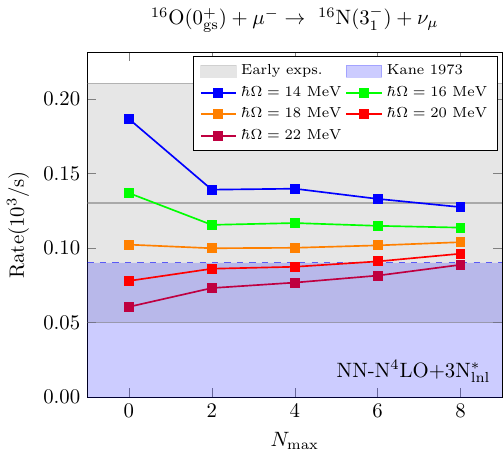}
    \end{subfigure}
    \caption{The dependence of selected OMC rates on the harmonic-oscillator frequency. The horizontal bands show the measured counterparts \cite{Deutsch1968, Measday2001, Abe2016,Kane1973}. `Early exps.' refers to the collection of measurements before 1973 from Ref. \cite{Measday2001}. The dashed line refers to an upper limit.}
    \label{fig:hw_dependence}
\end{figure*}

To extrapolate converged OMC rates, we study the convergence of the rates in terms of $N_{\rm max}$ and the HO frequency $\hbar\Omega$. We show examples of rates to different nuclear states obtained with the N$^4$LO+3N$^*_{\rm lnl}$ interaction as functions of $\hbar\Omega$ and $N_{\rm max}$ in Fig. \ref{fig:hw_dependence}. Available measured counterparts \cite{Deutsch1968, Measday2001, Abe2016,Kane1973} are shown as horizontal bands. Results for the other two interactions are very similar. We see that with increasing $N_{\rm max}$ the dependence on the HO frequency gets weaker, though the convergence patterns are generally different than for the ground-state energies in Fig. \ref{fig:ground_state_energies}. In ideal cases, like for the transitions to $^{12}$B($1^+_{\rm gs}$) and $^{16}$N($3^-_1$) in Fig. \ref{fig:hw_dependence}, the rates are constrained from above and below, depending on the frequency, and we can choose an optimal frequency between them. In other cases, like the transition to the $^6$He($0^+_{\rm gs}$), the convergence in terms of the frequency is less clear and we choose the frequency with the fastest convergence (in the case of $^6$He($0^+_{\rm gs}$) $\hbar\Omega=12$ MeV).

To further probe the quality of our calculations, we study the correlations between muon capture rates and other observables that are expected to have a similar operator structure. In particular, the transitions from $^6$Li($1^+_{\rm gs}$) to $^6$He($0^+_{\rm gs}$) and $^{12}$C($0^+_{\rm gs}$) to $^{12}$B($1^+_{\rm gs}$) with $\Delta J=1$ are driven by the $\bm{\sigma \tau}$ operator and can be related to Gamow-Teller $\beta$ decays and magnetic dipole (M1) $\gamma$ decays of the isobaric analog states (IAS). These processes operate at different momentum-exchange regimes and have different energy-dependencies, but strong correlations between spin-dominated processes at different momentum regimes have been found before in the context of $0\nu\beta\beta$ decay \cite{Shimizu2018,Romeo2021,Yao22,Jokiniemi2023b,Jokiniemi2023c}.

  \begin{figure*}
    \centering  
    \begin{subfigure}[b]{0.42\linewidth}
    \includegraphics[width=\linewidth]{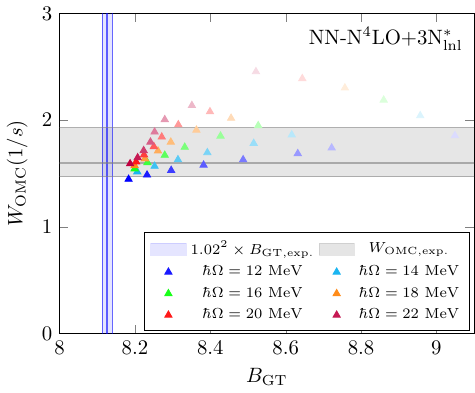}
    \end{subfigure}
    \begin{subfigure}[b]{0.42\linewidth}
    \includegraphics[width=\linewidth]{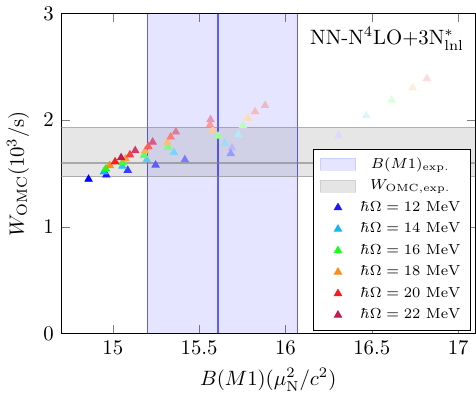}
    \end{subfigure}
    \caption{Muon-capture rate for $^6$Li($1^+_{\rm gs}$)+$\mu^-$ $\rightarrow$ $^6$He($0^+_{\rm gs}$)+$\nu_{\mu}$ vs. Gamow-Teller $\beta$-decay strength for $^6$He($0^+_{\rm gs}$) $\rightarrow$ $^6$Li($1^+_{\rm gs}$)+$e^-$+$\bar{\nu}_e$ (left) and M1 $\gamma$-decay strength for $^6$Li($0^+_1,T=1$) $\rightarrow$ $^6$Li($1^+_{\rm gs}$)+$\gamma$ (right) with different HO frequencies. The results are shown for $N_{\rm max}=0\dots14$; $N_{\rm max}$ increases as symbols become more opaque.}  
\label{fig:6Li_OMC_vs_BGT_and_M1}    
\end{figure*}

\begin{figure*}
    \centering
    \begin{subfigure}[b]{0.42\linewidth}
    \includegraphics[width=\linewidth]{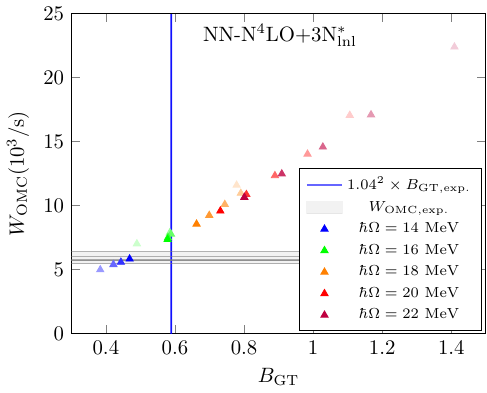}
    \end{subfigure}
    \begin{subfigure}[b]{0.42\linewidth}
     \includegraphics[width=\linewidth]{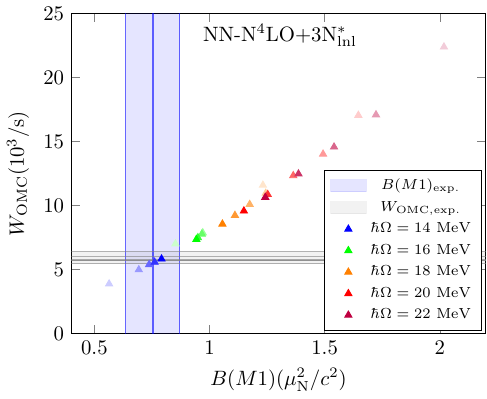}
     \end{subfigure}
    \caption{Muon-capture rate for $^{12}$C($0^+_{\rm gs}$)+$\mu^-$ $\rightarrow$ $^{12}$B($1^+_{\rm gs}$)+$\nu_{\mu}$ vs. GT $\beta$-decay strength for $^{12}$B($1^+_{\rm gs}$) $\rightarrow$ $^{12}$C($0^+_{\rm gs}$)+$e^-$+$\bar{\nu}_e$ (left) and M1 $\gamma$-decay strength for $^{12}$C($1^+_1,T=1$) $\rightarrow$ $^{12}$C($0^+_{\rm gs}$)+$\gamma$ (right) with different HO frequencies. The results are shown for $N_{\rm max}=0\dots8$; $N_{\rm max}$ increases as symbols become more opaque.}
    \label{fig:12C_OMC_vs_BGT_and_M1}
\end{figure*}

In Fig. \ref{fig:6Li_OMC_vs_BGT_and_M1} we compare the muon capture $^6$Li($1^+_{\rm gs}$)+$\mu^-$ $\rightarrow$ $^6$He($0^+_{\rm gs}$)+$\nu_{\mu}$ with the corresponding Gamow-Teller (GT) $\beta$ decay $^6$He($0^+_{\rm gs}$) $\rightarrow$ $^6$Li($1^+_{\rm gs}$)+$e^-$+$\bar{\nu}_e$ (left) and magnetic dipole (M1) $\gamma$ decay $^6$Li($1^+_{\rm gs}$) $\rightarrow$ $^6$Li($0^+_1,T=1$)+$e^-$+$\gamma$. We plot the GT strength
\begin{equation}
    B_{\rm GT}=\frac{g_{\rm A}^2}{2J_i+1}|\bra{\Psi_f}|\mathcal{O}_{\rm GT}\ket{\Psi_i}|^2\;,
\end{equation}
where $\mathcal{O}_{\rm GT}=\sum_{s=1}^A\boldsymbol{\sigma_s}\tau_+^s$, and the M1 transition strength
\begin{equation}
    B(M1)=\frac{1}{2J_i+1}|\bra{\Psi_f}|\mathcal{O}_{M1}|\ket{\Psi_i}|^2\;,
\end{equation}
where $\mathcal{O}_{M1}=\mu_N\sqrt{\frac{3}{4\pi}}\sum_{i=1}^A(g_i^l\boldsymbol{\ell}_i+\tfrac12 g_i^s\boldsymbol{\sigma}_i)$, against the corresponding OMC rate. All the observables are obtained without the inclusion of two-body currents as we are only interested in gross features here. We show the measured values \cite{NNDC, Deutsch1968, Friman-Gayer2021} for each process for comparison. To take the known effect of the two-body currents into account, we have multiplied the measured GT strength \cite{NNDC} by a factor $1.02^2$---it has been shown that the omission of two-body currents results in overestimation of the GT matrix element by $\sim 2\%$ \cite{Gysbers2019}. As expected given the similar operator structures, the GT and M1 strengths show good correlations with the muon-capture rate. Different HO frequencies (shown as different colors) give slightly different correlation lines, indicating that the GT and M1 observables have different HO-frequency dependencies than muon capture. In the left panel, we see that with increasing $N_{\rm max}$ (increasing color gradient) we reach the measured GT strength and slightly underestimate the muon-capture rate. In the right panel, one can see that we slightly underestimate the M1 strengths, but the inclusion of two-body currents is expected to increase the B(M1) values by some $5-10$\% \cite{Friman-Gayer2021}.

In Fig. \ref{fig:12C_OMC_vs_BGT_and_M1} we show similar results for the $^{12}$C: we compare the capture rate from $^{12}$C($0^+_{\rm gs}$) to $^{12}$B($1^+_{\rm gs}$) with the GT $\beta$ decay of $^{12}$B($1^+_{\rm gs}$) (left panel) and with the M1 $\gamma$-decay of $^{12}$C($1^+$;1)---the IAS of $^{12}$B($1^+_{\rm gs}$) (right panel). The measured values \cite{NNDC, Measday2001,Abe2016} are shown for comparison. Again, we multiply the measured GT strength by $1.04^2$ to correct for the effect of the omitted two-body currents: for $^{16}$O the two-body currents are found to reduce the GT NMEs by $\sim 4\%$ \cite{Gysbers2019}. We see that both the observables are strongly correlated with the capture rate and in this case the correlations do not depend on the HO frequency. With the optimal frequency, $\hbar\Omega=16$ MeV (see Fig. \ref{fig:hw_dependence}), we reproduce the measured GT strength very well and slightly overestimate the M1 decay and the muon capture rate. Including the two-body currents into the M1 decay calculation could change the situation.

\subsection{Partial Muon-Capture Rates to Low-Lying States in $^{6}$He, $^{12}$B and $^{16}$N}

\begin{figure}
    \centering
\includegraphics{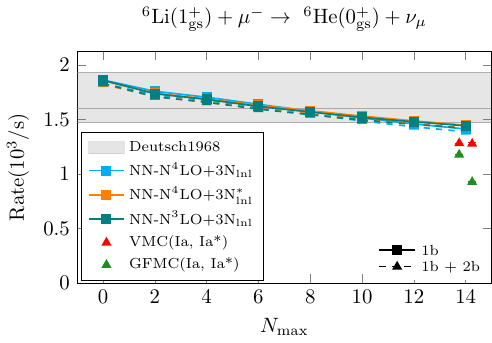}
    \caption{Capture rate to the ground state of $^{6}$He obtained with different interactions with $\hbar\Omega=12$ MeV as a function of $N_{\rm max}$. The dashed (solid) lines include one- and two-body currents (one-body currents only). The triangles denote the variational (red) and Green's function (green) Monte Carlo results \cite{King2022} including both one- and two-body currents obtained with two different Norfolk two- and three-body interactions (Ia, Ia$^*$). The horizontal band shows the measured rate \cite{Deutsch1968}.}
    \label{fig:6Li_gs}
\end{figure}

We collect the partial muon-capture rates to the ground state of $^6$He, obtained with the three different interactions, in Fig. \ref{fig:6Li_gs}. For each interaction, we choose the frequency $\hbar\Omega=12$ MeV. The interaction dependence of the transition is very moderate: all the employed interactions agree with each other within a few \%. While the agreement with the experiment \cite{Deutsch1968} is reasonable, the convergence of the rate is rather slow and all the interactions underestimate the experiment by some $10\%-20\%$. This is likely due to the cluster-structure of the involved nuclei, which is not fully captured by the NCSM without continuum. We also compare the results obtained with one-body currents only (solid lines) against those obtained by adding the effect of axial-vector two-body currents (dashed lines) via the Fermi-gas model. For this light case, the two-body currents reduce the capture rates only by $\lesssim 2\%$.

\begin{figure}
    \centering   
    \includegraphics[width=\linewidth]{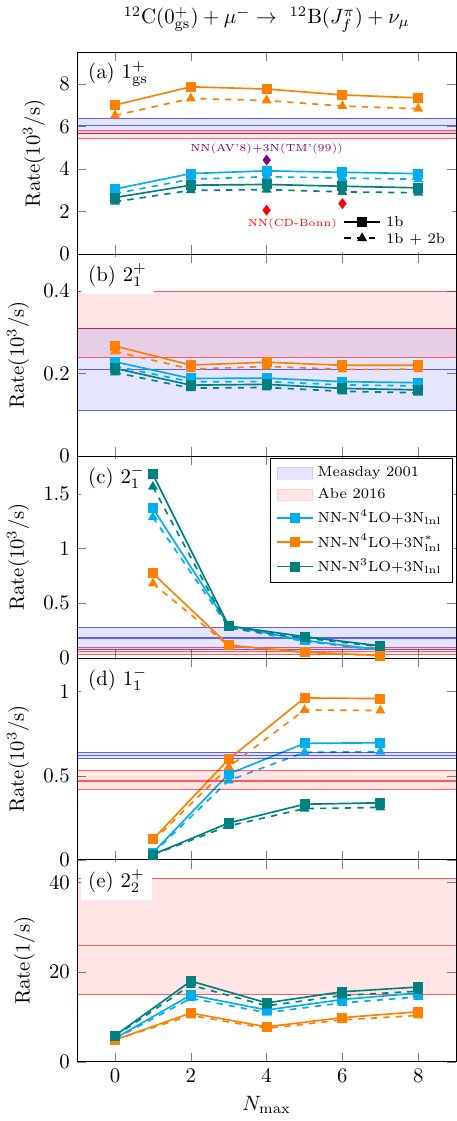}
    \caption{Muon-capture rates to different $J^{\pi}_f$ final states in $^{12}$B obtained with different chiral interactions as functions of $N_{\rm max}$. The triangles show the NCSM results from Ref. \cite{Hayes2003}, obtained with CD-Bonn two-nucleon force only (red) and with AV'8 two-nucleon and TM'(99) three-nucleon forces (purple). The rates are obtained with HO frequencies with best convergence properties. The horizontal bands denote measured rates with uncertainties \cite{Measday2001,Abe2016}.}
    \label{fig:Rates_12B}
\end{figure}

\begin{figure}
    \centering
    \includegraphics[width=\linewidth]{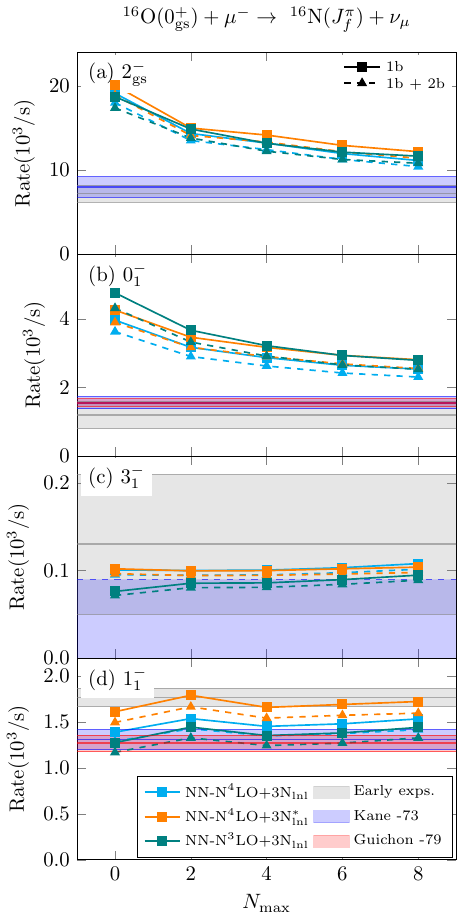}
    \caption{Capture rates to different $J^{\pi}_f$ final states in $^{16}$N. The rates are obtained with HO frequencies with best convergence properties. The blue and red bands are the measured rate with uncertainties from \cite{Kane1973,Guichon1979}. The gray band shows the combined results of early experiments taken from \cite{Measday2001}.}
    \label{fig:Rates_16N}
\end{figure}

We extrapolate the final results by fitting an exponential to the rates shown in Fig. \ref{fig:6Li_gs}, and estimate the uncertainty by varying the points taken into account in the fitting. The extrapolated rates are shown in Table \ref{tab:rates}. Within the estimated accuracy, all three interactions are consistent with each other and underestimate the experiment by some 20\%. Our results, within the uncertainty, are in good agreement with the values obtained with quantum Monte Carlo \emph{ab initio} methods with Norfolk two-nucleon and three-nucleon local chiral interactions \cite{King2022}. The corresponding variational Monte Carlo (VMC) method results with one- and two-body currents were $1282\pm2$ 1/s with the NV2+3-1a model and $1177\pm2$ 1/s with the NV2+3-1a$^*$ model, both underestimating the measured rate by some $20\%-25\%$. In the Green's function Monte Carlo method, the rates are reduced to $1277\pm10$ 1/s (NV2+3-1a model) and $926\pm8$ 1/s (NV2+3-1a$^*$ model), increasing the discrepancy with the experimental counterpart.

\begin{table*}[t]
    \centering
    \caption{The partial capture rates for the different transitions obtained with different interactions with or without the approximated two-body currents. The measured rates are given in the last column.}
    \begin{ruledtabular}
    \begin{tabular}{llcccc|c}
Final state& Interaction &$W({\rm 1b})(1/{\rm s})$ &$W({\rm 1b+2b})(1/{\rm s})$ &Expt. (1/s)\\
\hline
$\rm ^6He(0^+_{gs})$ &NN-N$^3$LO+3N$_{\rm lnl}$ &1300(100) &1300(100) &$1600\substack{+330 \\ -129}$ \cite{Deutsch1968}\\
&NN-N$^4$LO+3N$_{\rm lnl}$  &1300(100) &1300(100)\\
&NN-N$^4$LO+3N$^*_{\rm lnl}$  &1300(100) &1300(100)\\ \hline
$\rm ^{12}B(1^+_{gs})$ &NN-N$^3$LO+3N$_{\rm lnl}$  &3100(700) &2900(700) &$6040\pm350$ \cite{Measday2001}\\
&NN-N$^4$LO+3N$_{\rm lnl}$  &3800(700) &3500(700) &$5680\substack{+140 \\ -230}$ \cite{Abe2016}\\
&NN-N$^4$LO+3N$^*_{\rm lnl}$  &7400(700) &6800(700)\\ \hline
$\rm ^{12}B(2^+_1)$ &NN-N$^3$LO+3N$_{\rm lnl}$ &160(10) &150(10) &$210\pm100$ \cite{Measday2001}\\
&NN-N$^4$LO+3N$_{\rm lnl}$  &180(10) &170(10) &$310\substack{+90 \\ -70}$ \cite{Abe2016}\\
&NN-N$^4$LO+3N$^*_{\rm lnl}$ &220(10) &210(10)\\ \hline
$\rm ^{12}B(2^+_2)$ &NN-N$^3$LO+3N$_{\rm lnl}$ &17(5) &16(5) &$26\substack{+15 \\ -11}$ \cite{Abe2016}\\
&NN-N$^4$LO+3N$_{\rm lnl}$  &15(5) &14(5)\\
&NN-N$^4$LO+3N$^*_{\rm lnl}$ &13(5) &12(5)\\ \hline
$\rm ^{12}B(1^-_1)$ &NN-N$^3$LO+3N$_{\rm lnl}$  &340(50)  &310(50) &$620\pm20$ \cite{Measday2001}\\
&NN-N$^4$LO+3N$_{\rm lnl}$  &700(50) &640(50) &$470\substack{+60 \\ -50}$ \cite{Abe2016}\\
&NN-N$^4$LO+3N$^*_{\rm lnl}$ &960(50)  &890(50)\\ \hline
$\rm ^{12}B(2^-_1)$ &NN-N$^3$LO+3N$_{\rm lnl}$  &60(50) &60(50)  &$180\pm 100$ \cite{Measday2001}\\
&NN-N$^4$LO+3N$_{\rm lnl}$  &50(40)  &50(40) &$60\substack{+40\\-30}$\cite{Abe2016}\\
&NN-N$^4$LO+3N$^*_{\rm lnl}$  &20(20) &20(20)\\ \hline
$\rm ^{16}N(2^-_{gs})$ &NN-N$^3$LO+3N$_{\rm lnl}$ &11100(1000) &10300(1000) &$7200\pm1000$ \cite{Measday2001}\\
&NN-N$^4$LO+3N$_{\rm lnl}$  &10200(1000) &9400(1000) &$8000\pm1200$ \cite{Kane1973}\\
&NN-N$^4$LO+3N$^*_{\rm lnl}$  &11200(1000) &10400(1000)\\ \hline
$\rm ^{16}N(0^-_1)$ &NN-N$^3$LO+3N$_{\rm lnl}$  &2600(200) &2400(200) &$1200\pm400$ \cite{Measday2001}\\
&NN-N$^4$LO+3N$_{\rm lnl}$ &2400(200) &2200(200) &$1560\pm180$ \cite{Kane1973}\\
&NN-N$^4$LO+3N$^*_{\rm lnl}$  &2600(200) &2400(200) &$1560\pm110$ \cite{Guichon1979}\\ \hline
$\rm ^{16}N(3^-_1)$ &NN-N$^3$LO+3N$_{\rm lnl}$ &110(20) &100(20) &$130\pm80$ \cite{Measday2001}\\
&NN-N$^4$LO+3N$_{\rm lnl}$  &110(20) &100(20) &$\leq90$ \cite{Kane1973}\\
&NN-N$^4$LO+3N$^*_{\rm lnl}$  &110(20) &100(20)\\ \hline
$\rm ^{16}N(1^-_1)$ &NN-N$^3$LO+3N$_{\rm lnl}$ &1600(100) &1500(100) &$1770\pm100$ \cite{Measday2001}\\
&NN-N$^4$LO+3N$_{\rm lnl}$  &1700(100) &1600(100) &$1310\pm110$ \cite{Kane1973}\\
&NN-N$^4$LO+3N$^*_{\rm lnl}$ &1800(100) &1700(100) &$1270\pm90$ \cite{Guichon1979}\\ 
    \end{tabular}
    \end{ruledtabular}
    \label{tab:rates}
\end{table*}

The partial capture rates to the low-lying states in $^{12}$B obtained with the three different interactions are collected in Fig. \ref{fig:Rates_12B} and the extrapolated rates are shown in Table \ref{tab:rates}. For each interaction, we show the results obtained with the HO frequency giving the best convergence (see Fig. \ref{fig:hw_dependence}). We note that in general the results are in reasonable agreement with the measured capture rates with uncertainties, shown as blue \cite{Measday2001} and red \cite{Abe2016} bands. The effect of two-body currents varies between $5\%-10\%$ for each transition.

The NN-N$^3$LO+3N$_{\rm lnl}$ and NN-N$^4$LO+3N$_{\rm lnl}$ interactions underestimate the measured rates to the $1^+$ ground-state of $^{12}$B by some $30\%-45\%$, while the N$^4$LO+3N$^*_{\rm lnl}$ interaction slightly overestimates them though being consistent within the error bars. As a Gamow-Teller--type transition, the capture rate to the $1^+$ ground-state of $^{12}$B is sensitive to the spin-orbit interaction. Hence, including the sub-leading $E_7$ spin-orbit term \cite{Girlanda2011} into the 3N-part of the NN-N$^4$LO+3N$_{\rm lnl}$ interaction has a significant effect on the capture rate. The fact that this interaction improves the agreement with the experiments supports the addition of the extra $E_7$ term.

The obtained partial capture rate to the  ground state of $^{12}$B shown in Fig. \ref{fig:Rates_12B} can be compared with the earlier NCSM results \cite{Hayes2003}, obtained with less realistic interactions. We notice that our results are above the rate $2.38\times10^3$ 1/s \cite{Hayes2003} obtained with the CD-Bonn NN interaction \cite{Machleidt1996} only at $N_{\rm max}=6$. On the other hand, our results are comparable to that obtained with the value $4.43\times10^3$ 1/s obtained with AV8' NN interaction \cite{Pudliner1997} accompanied by the TM'(99) 3N interaction \cite{Coon2001} at $N_{\rm max}=4$. 

For the $2^+$ and $2^-$ states in $^{12}$B (see panels (b), (c) and (e) in Fig. \ref{fig:Rates_12B}), the interaction dependence is less significant. However, for the lowest $2^+$ state, the NN-N$^4$LO+3N$^*_{\rm lnl}$ interaction with the $E_7$ term again brings the rate up, in better agreement with the measured rates, compared to the other two interactions. On the other hand, the NN-N$^4$LO+3N$^*_{\rm lnl}$ interaction underestimates the measured rate to the $2^+_2$ state, while the other two interaction reach the lower limit by increasing $N_{\rm max}$. Nonetheless, it should be noted that the rate to this state is very low, one to two orders of magnitude lower than for the other states of interest, so this particular transition should not be given undue weight. 

The capture rate to the $1^-_1$ state in $^{12}$B is particularly sensitive to the chiral interaction: the rate obtained with the NN-N$^3$LO+3N$_{\rm lnl}$ interaction is half of that obtained with the NN-N$^4$LO+3N$_{\rm lnl}$ and only one third of that obtained with the NN-N$^4$LO+3N$^*_{\rm lnl}$ interaction. Also the two measured rates disagree with each other, so it is hard to draw definite conclusions about this transition. However, the NN-N$^4$LO+3N$_{\rm lnl}$ is in best agreement with these experiments.

The rates to different final states in $^{16}$N, obtained with the optimal HO frequencies, are shown in Fig. \ref{fig:Rates_16N}, and the extrapolated rates are summarized in Table \ref{tab:rates}. One can see that the rates are generally in good agreement with the measured rates \cite{Measday2001,Kane1973,Guichon1979}, ranging from $\sim 100$ 1/s to $10^4$ 1/s depending on the final state. Contrary to the $^{12}$C case, the interaction-dependence is quite mild: $\sim 25\%$ for the transition rate to the $1^-_1$ state (panel (d)) and $5\%-10\%$ for the rest of the transitions (panels (a)-(c)). The effect of two-body currents is similar to the $^{12}$C case, between $5\%-10\%$.

\begin{figure*}
    \centering
    \begin{subfigure}[b]{0.4\linewidth}
\includegraphics[height=0.9\linewidth]{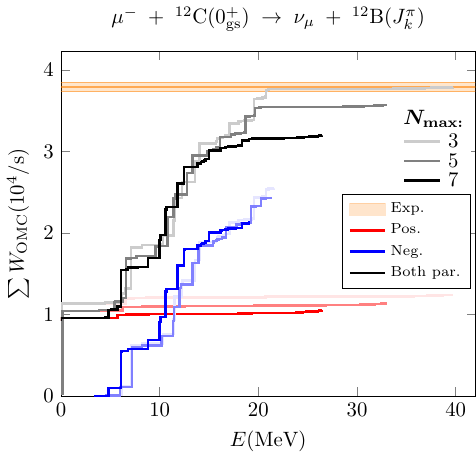}
    \end{subfigure}
     \begin{subfigure}[b]{0.4\linewidth}
    \includegraphics[height=0.9\linewidth]{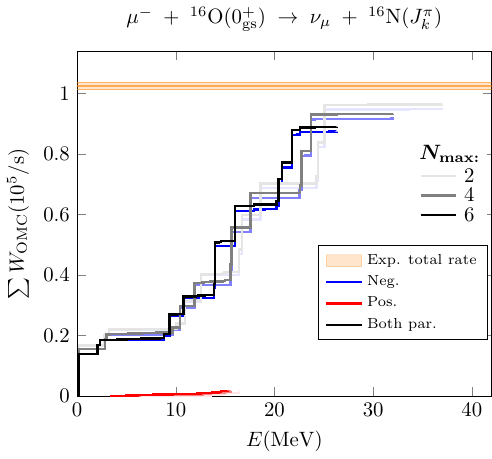}
    \end{subfigure}
    \caption{The summed OMC rates to positive- (red lines) and negative-parity (blue lines) final states in $^{12}$B (left) and $^{16}$N (right). The rates are computed with the NN-N$^4$LO+3N$^*_{\rm lnl}$ interaction summing over several dozens of final states of each parity (see the text for details). Two-body currents are omitted here for simplicity. The color gradient of the lines denotes increasing $N_{\rm max}$. The orange band shows the experimental total rate \cite{Measday2001}.}
    \label{fig:total_rates}
\end{figure*}

Having a closer look at the rates, the convergence of the rates to the $2^-_{\rm gs}$ and $0^-_1$ state (panels (a) and (b) in Fig. \ref{fig:Rates_16N}) is very similar, and the NN-N$^3$LO+3N$_{\rm lnl}$ and NN-N$^4$LO+3N$^*_{\rm lnl}$ interactions give very similar rates, while the NN-N$^4$LO+3N$_{\rm lnl}$ interactions gives $\sim 10\%$ smaller rates. In both cases, we overestimate the measured rates; for the $2^-_{\rm gs}$ by some $10\%-20\%$ and for the $0^-_1$ state by some $40\%-50\%$. The reason for the overestimation is currently unclear and might become more apparent with the inclusion of exact two-body currents.

For the $3^-_1$ and $1^-_1$ states, the dependence on the $N_{\rm max}$ is very weak, and we extrapolate the final results in Table \ref{tab:rates} from the $N_{\rm max}$ results with the uncertainty coming from the variation of the $N_{\rm max}$ and $\hbar\Omega$. For the transition to the $3^-_1$ state, all the interactions are consistent with the early experiments \cite{Measday2001} and, after the inclusion of the two-body currents, consistent with the upper limit of Ref. \cite{Kane1973} within the error bars. For the $1^-_1$ state, the NN-N$^3$LO+3N$_{\rm lnl}$ is consistent with Refs. \cite{Kane1973,Guichon1979} within the error bars, while the other two interactions agree with the early experiments \cite{Measday2001} within the error bars.

\subsection{Total capture rates}

In order to study the overall performance of our theory formalism, we also approximate the
total muon-capture rates in the same nuclei by a simple summation over the final states. In Fig. \ref{fig:total_rates}, we show the total rate obtained by summing over all the computationally accessible final states in $^{12}$B. We restrict the number of both positive- (red) and negative-parity (blue) states to a maximum of 75 states for technical reasons. The color gradients of the lines denote increasing $N_{\rm max}$: the lightest (darkest) ones correspond to smallest (biggest) $N_{\rm max}$. The excitation energies of the negative-parity states are given with respect to the ground-state energy obtained with the preceding even $N_{\rm max}$. The black lines denote the sum of the rates to the positive- and negative-parity states. The orange band denotes the total capture rate taken from Ref. \cite{Measday2001}.

For $^{12}$C, the rates to positive-parity states are obtained with $N_{\rm max}=2,4,6$ and those for the negative-parity states with $N_{\rm max}=3,5,7$. We combine the rates to positive- and negative-parity states by summing $N_{\rm max}=2$ with the $N_{\rm max}=3$, $N_{\rm max}=4$ with the $N_{\rm max}=5$, and $N_{\rm max}=6$ with the $N_{\rm max}=7$. This is a natural way to combine the results, since the negative-parity states for $N_{\rm max}+1$ are built on top of the positive-parity states obtained with $N_{\rm max}$. Altogether, we include 75 states for each $N_{\rm max}$, except for $N_{\rm max}=7$ for which we only reach 52 converged states. Since the NCSM calculations for the $A=16$ systems are much heavier than for the $A=12$ systems, we cannot reach a large number of converged states with $N_{\rm max}=7$. Hence, here the rates to positive-parity states are obtained with $N_{\rm max}=1,3,5$, those to the negative-parity states with $N_{\rm max}=2,4,6$ and the summed rates are obtained by summing $N_{\rm max}=1$ with $N_{\rm max}=2$, $N_{\rm max}=3$ with $N_{\rm max}=4$ etc. Since these $N_{\max}$ pairs are not consistent with each others, we use the measured energies \cite{NNDC} to adjust the difference of the energies of the lowest negative- and positive-parity final states. In this case, we obtained 55 converged states for $N_{\rm max}=1,3,5$; 75 states for $N_{\rm max}=2,4$; and 64 states for $N_{\rm max}=6$.

Fig. \ref{fig:total_rates} shows that summing up the rates over the computed states we are able to reproduce most of the measured total capture rates and underestimate them only by some 15\% in both cases. The discrepancy can be understood looking at the cumulative sums as functions of energy: the higher the $N_{\rm max}$ the lower the excitation energies we can reach. With large $N_{\rm max}$, the density of states becomes high and, for technical reasons, we cannot reach highly excited states potentially contributing substantially to the summed rate, suggested by the lower-$N_{\rm max}$ distributions. This is also supported by experimental data: in Ref. \cite{Hashim2018,Hashim2021}
it was seen that in the case of $^{100}$Nb, the measured OMC strength function reaches excitation energies up to 50 MeV, and in particular there is high-energy giant-resonance at around 30 MeV. The situation in these lighter nuclei could be similar. It is worth noting that the summation method we use here is not the most efficient way to obtain the total capture rates. In a future work, we plan to revisit the total muon-capture rates using the Lanczos strength function method that likely allows us to capture the total muon-capture strength with less computing power.

\section{Summary and outlook}
\label{sec:summary}
We study muon capture on light nuclei $^6$Li, $^{12}$C and $^{16}$O using no-core shell-model with three different chiral interactions including three-body forces. We solve the bound-muon wave functions from the Dirac equations to take into account relativistic corrections and the finite size of the nucleus. We remove the spurious center-of-mass motion from the operators by introducing translationally invariant operators. Furthermore, we investigate the effect of axial-vector two-body currents via effective one-body currents normal-ordered with respect to a Fermi-gas reference state.

We test the validity of the nuclear wave functions by comparing the calculated ground-state and excited-state energies against the experimental counterparts and find a good agreement. We study the convergence of the muon-capture rates in terms of the harmonic-oscillator basis size and frequency and extrapolate the results from the convergence patters. We find in general good agreement with the measured muon-capture rates as well as with earlier \emph{ab initio} calculations obtained with different chiral interactions. We observe sensitivity of some of the rates to the sub-leading chiral three-nucleon interaction terms. The Fermi-gas-approximated two-body currents are found to reduce the capture rates by $\sim2\%$ in $^6$Li and by $< 10\%$ in $^{12}$C and $^{16}$O. Including exact two-body currents is left for a future work. We also estimate total muon-capture rates in $^{12}$C and $^{16}$O by summing the partial capture rates over all the computationally accessible final states in $^{12}$B and $^{16}$N. In both cases, we obtain roughly $85\%$ of the measured total rate, the underestimation being likely due to the restricted number of final states. A better estimation of the total capture rates, using the Lanczos strength-function method is underway. 

The present study serves as one step closer to better understanding of the weak interaction at finite momentum transfer---relevant for neutrinoless double-beta decay. To draw further conclusions, both theoretical and experimental efforts are called for. From the theory side, inclusion of exact two-body currents and continuum states would help us improve the calculations. Eventually, one would also need to derive robust uncertainties for the capture rates, taking into account the errors coming from both the truncation of the $\chi$EFT expansion and the many-body method. It would also be beneficial to revisit the measurements, especially for the transitions for which the earlier measurements disagree, and to extend the measurements to other nuclei accessible for the \emph{ab initio} techniques. Nevertheless, the good description of the muon-capture processes also motivates further studies of similar processes involving the same nuclei. For example, both $^{12}$C and $^{16}$O are interesting candidates for neutrino-scattering experiments \cite{Balantekin2022}.

\begin{acknowledgments}
This work was supported by Arthur B. McDonald Canadian Astroparticle Physics Research Institute, by the NSERC Grant No. SAPIN-2022-00019, by the U.S. Department of Energy, Office of Science, Office of Nuclear Physics, under Work Proposals No. SCW0498, and by the Academy of Finland, Grant Nos. 314733 and 345869. KK receives partial support through the Scientific Discovery through Advanced Computing (SciDAC) program funded by U.S. Department of Energy, Office of Science, Advanced Scientific Computing Research and Nuclear Physics. TRIUMF receives federal funding via a contribution agreement with the National Research Council of Canada. This work was prepared in part by LLNL under Contract No. DE-AC52-07NA27344. Computing support came from an INCITE Award on the Summit and Frontier supercomputers of the Oak Ridge Leadership Computing Facility (OLCF) at ORNL and from the Digital Research Alliance of Canada.
\end{acknowledgments}

\bibliography{refs}

\end{document}